\documentstyle[12pt]{article}

\newlength{\dinwidth}
\newlength{\dinmargin}
\def\lapproxeq{\lower .7ex\hbox{$\;\stackrel{\textstyle <}{\sim}\;$}}
\def\gapproxeq{\lower .7ex\hbox{$\;\stackrel{\textstyle >}{\sim}\;$}}

\begin{document}

\newfont{\sevenrm}{cmr7}
\newfont{\teni}{cmmi10}
\newfont{\seveni}{cmmi7}
\newfont{\sevensy}{cmsy7}
\newfont{\fiverm}{cmr5}
\newfont{\fivei}{cmmi5}
\newfont{\fivesy}{cmsy5}

\def\tenpoint{\normalbaselineskip=12pt
\abovedisplayskip 12pt plus 3pt minus 9pt
\belowdisplayskip 12pt plus 3pt minus 9pt
\abovedisplayshortskip 0pt plus 3pt
\belowdisplayshortskip 7pt plus 3pt minus 4pt
\smallskipamount=3pt plus1pt minus1pt
\medskipamount=6pt plus2pt minus2pt
\bigskipamount=12pt plus4pt minus 4pt
\def\rm{\fam0\tenrm}          \def\it{\fam\itfam\tenit}%
\def\sl{\fam\slfam\tensl}      \def\bf{\fam\bffam\tenbf}%
\def\smc{\tensmc}               \def\mit{\fam 1}%
\def\cal{\fam 2}%
\textfont0=\tenrm      \scriptfont0=\sevenrm    \scriptscriptfont0=\fiverm
\textfont1=\teni       \scriptfont1=\seveni     \scriptscriptfont1=\fivei
\textfont2=\tensy      \scriptfont2=\sevensy    \scriptscriptfont2=\fivesy
\textfont3=\tenex      \scriptfont3=\tenex      \scriptscriptfont3=\tenex
\normalbaselines\rm}

\titlepage
\begin{flushright}
RAL-95-021 \\
DTP/95/14 \\
February 1995
\end{flushright}
\vskip 2.cm

\begin{center}
{\Large\bf Pinning down the Glue in the Proton}
\vskip 1.cm
{\large A.D.~Martin} and {\large W.J.~Stirling}
\vskip .2cm
{\it Department of Physics, University of Durham \\
Durham DH1 3LE, England }\\
\vskip .4cm
and
\vskip   .4cm
{\large R.G.~Roberts}
\vskip .2cm
{\it
Rutherford Appleton Laboratory,  \\
Chilton, Didcot  OX11 0QX, England
} \\
\end{center}
\vskip 1cm

\begin{abstract}
The latest measurements of $F_2$ at HERA allow for a {\it combination}
of gluon and sea quark distributions at small $x$ that is significantly
 different from those of existing parton sets.  We perform a new global fit to
deep-inelastic and related data. We find  a gluon distribution which is
larger for $x \lapproxeq 0.01$, and  smaller
for $x \sim 0.1$, and a flatter input sea quark distribution than those
obtained in our most recent global analysis.
The new fit also gives $\alpha_s(M_Z^2) = 0.114$. We study other
experimental information available for the gluon including, in
particular, the constraints coming from fixed-target and collider
prompt $\gamma$ production data.
\end{abstract}

\newpage

The advent of the high-energy electron-proton collider, HERA, has enabled the
partonic structure of the proton to be investigated in the small $x$ regime, $x
\lapproxeq 10^{-3}$.  The first measurements showed that the proton structure
function $F_2(x,Q^2)$ rose dramatically with decreasing $x$.  When these HERA
data were included in global analyses of deep-inelastic and
related data, the sea
quark and gluon distributions were typically found to have the small $x$
behaviour $xS, xg \sim x^{-0.3}$.  To be precise, in the most recent
MRS analysis
\cite{MRSA} the sea quark and gluon distributions are
parametrized in the form
\begin{eqnarray}
xS & =& A_Sx^{-\lambda_S} (1 + \epsilon_S\sqrt{x}
                           + \gamma_Sx) (1 - x)^{\eta_S} \\
xg & = & A_g x^{-\lambda_g} (1 + \epsilon_g \sqrt{x} + \gamma_gx)(1-x)^{\eta_g}
\end{eqnarray}
with $\epsilon_g \equiv 0$ and $\lambda_S \equiv \lambda_g = \lambda$.  At that
time there were insufficient experimental constraints on the gluon to justify
the
introduction of the parameter $\epsilon_g$ in $xg$, or to determine the
exponent
$\lambda_g$ independent of that of the sea-quark distribution $S$.  The global
fit
\cite{MRSA} gave $\lambda = 0.3 \pm 0.1$ where the value was strongly
correlated
to $\epsilon_S$. An independent global analysis by the CTEQ
collaboration \cite{CTEQ3} gave similar results with a common
 $\lambda$ of $0.286$.
Thus the small-$x$ behaviour of the sea-quark and gluon distributions
are closely linked --  motivated either by BFKL gluon dynamics with the sea
quarks driven by the $g \to q \bar q$  transition, or   by GLAP
evolution from a low starting $Q^2$ scale.

The most recent measurements of $F_2(x,Q^2)$ presented by the ZEUS \cite{ZEUS}
and H1 \cite{H1} collaborations have more data points at low $x$ and
improved precision compared to those presented earlier.
The new measurements are shown in Fig.~1, together with their
description by MRS(A) \cite{MRSA} partons (dash-dot curves). At very small $x$
we see that the MRS(A) values lie above the new data. The same is true
of the new CTEQ3 partons \cite{CTEQ3} and, to a
lesser extent, of the latest set of
GRV(94) partons \cite{GRV} (dotted curves).   We therefore repeat
the global analysis of Ref.~\cite{MRSA} with the new HERA data included.
We allow the normalization of the ZEUS \cite{ZEUS} and H1 \cite{H1}
data sets to vary within their quoted uncertainties ($\pm 3.5\%$ and
$\pm 4.5\%$ respectively).
The resulting partons\footnote{The {\tt FORTRAN} code for the A$'$
and G sets are available by
electronic mail from W.J.Stirling@durham.ac.uk}, which we denote A$'$,
correspond to the following parametrization of the starting distributions at
$Q^2_0 = 4$~GeV$^2$
\begin{eqnarray}
xu_v & = & A_u \; x^{0.559} (1 - 0.54 \sqrt{x} + 4.65 x)(1-x)^{3.96} \\
xd_v & = & A_d\; x^{0.335} (1 + 6.80 \sqrt{x} + 1.93 x)(1-x)^{4.46} \\
xS & = & 0.956\; x^{-0.17} (1 - 2.55 \sqrt{x} + 11.2 x)(1-x)^{9.63} \\
xg & = & A_g\; x^{-0.17} (1 - 1.90 \sqrt{x} +4.07 x)(1-x)^{5.33} .
\end{eqnarray}
The sea $S=2(\bar u + \bar d + \bar s + \bar c)$ has a
 flavour structure similar to that of MRS(A), with
parameters $A_{\Delta} = 0.045$,
$\gamma_{\Delta} = 49.6$, $\eta_{\Delta} = 0.3$,
$\delta = 0.02$  and $m^2_c = 2.7$~GeV$^2$ in the notation of
Eqs.~(7) and (8) of  Ref.~\cite{MRSA}.
The QCD scale parameter is
found to be $\Lambda_{\overline{\rm MS}}(n_f=4) = 231$~MeV, which
corresponds to $\alpha_s(M^2_Z) = 0.113$.  The parameters $A_u$ and
$A_d$ are determined by the flavour sum rules,
giving 2.26 and 0.279 respectively,
and $A_g = 1.94$ is determined by
the momentum sum rule.

The A and A$'$ partons differ only in the small-$x$ region,
 since the data have changed only in this region. Indeed the main effect
of the new HERA data is to change the small-$x$ behaviour  $\lambda
\equiv \lambda_S = \lambda_g = 0.3$ of MRS(A) to $\lambda = 0.17$ of
MRS(A$'$). Besides the change in $\lambda$, in the A$'$ analysis
we allowed $\epsilon_g$ to be a free parameter, which also helped
to significantly improve the fit.
The dashed curves of Fig.~1 show how the description  of the
small-$x$ data is improved. However it is evident that the HERA data
suggest stronger scaling violations than the new A$'$ fit.
Now, for $x \lapproxeq 0.01$, the slope
\begin{equation}
\frac{\partial F_2(x,Q^2)}{\partial \log Q^2} \simeq
\frac{\alpha_s(Q^2)}{\pi} \sum_q e^2_q \int^1_x \frac{dy}{y} \left( \frac{x}{y}
\right) P_{qg} \left( \frac{x}{y} \right) yg(y,Q^2) ,
\label{slope}
\end{equation}
is virtually a direct indicator of the magnitude of the gluon
distribution. Indeed both the ZEUS and H1 collaborations \cite{HERAG}
have used their respective data to determine the slope, and hence the
gluon distribution, and find values for the gluon in excess of
MRS(A$'$),
more consistent in fact with the steeply rising gluon distributions of,
for example, MRS(D$_-'$) \cite{MRSD}. On the one hand the new
A$'$ fit to $F_2$ gives a flatter sea (and gluon) and yet on the
other hand the slope $\partial F_2(x,Q^2) / \partial \log Q^2$
seems to prefer a steeper gluon.

The situation is well illustrated by Fig.~2. The upper plot for
$\partial F_2(x,Q^2) / \partial \log Q^2$ is an indicator of the gluon,
see Eq.~(\ref{slope}), whereas the lower plot is dominated by the sea
quark distribution. The new HERA data show that  the dashed A$'$
curve, obtained assuming $\lambda_g = \lambda_S$, is too flat (steep) in
the upper (lower) plot. We conclude that
the small-$x$ data are now sufficiently precise
that their description  requires partons with   $\lambda_g \neq
\lambda_S$. We therefore repeat the global analysis with both
$\lambda_g$ and $\lambda_S$ free and find  at
$Q^2_0 = 4$~GeV$^2$
\begin{eqnarray}
xu_v & = & A_u\; x^{0.593} (1 - 0.76 \sqrt{x} + 4.20 x)(1-x)^{3.96} \\
xd_v & = & A_d\; x^{0.335} (1 + 8.63 \sqrt{x} + 0.32 x)(1-x)^{4.41} \\
xS & = & 1.74\; x^{-0.067} (1 - 3.45 \sqrt{x} + 10.3 x)(1-x)^{10.1} \\
xg & = &  A_g\; x^{-0.301} (1 - 4.14 \sqrt{x} +10.1 x)(1-x)^{6.06} ,
\label{gluon}
\end{eqnarray}
with $A_g = 1.51$; $A_{\Delta} = 0.043$ and
$\gamma_{\Delta} = 64.9$,  the
remaining parameters being unchanged from A$'$.
The QCD scale parameter increases to
\begin{equation}
\Lambda_{\overline{\rm MS}}(n_f=4) = 255 \, {\rm MeV} ,
\end{equation}
which corresponds to $\alpha_s(M^2_Z) = 0.114$. The resulting set of
partons$^1$ is denoted  by MRS(G), and gives rise  to the continuous curves
in Figs.~1 and 2. We see that the steeper gluon ($\lambda_g = 0.30$)
and the flatter sea ($\lambda_S = 0.07$) significantly improve the
description of the small-$x$ HERA data.
  Due to the negative $\epsilon_g \sqrt{x}$
term in $xg$, the effective value of $\lambda_g$ is larger than the
value shown.  To be precise, if we approximate Eq.~(\ref {gluon}) by
$xg=Ax^{-\bar\lambda}$ over the interval $10^{-4}<x<5\times 10^{-3}$
then we find $\bar\lambda=0.355$.
Note that the data shown in Fig.~2(b) are the result of extrapolating
measured data from higher values of $Q^2$. Even if future low $Q^2$
measurements continue to show a rise for $F_2$ at small $x$ the slopes
extracted from Fig.~2(a) would still allow a steep gluon solution.

The new HERA measurements of $F_2$ provide for the first time a reliable
estimate of the gluon at small $x$. We can immediately see the trends.
If it is indeed true that the HERA data prefer
an enhancement of the A$'$ gluon for
$x \sim 10^{-3}$ \cite{HERAG} then there are implications  for other
gluon-dominated reactions, in particular for prompt $\gamma$ production.
 The fixed-target prompt $\gamma$ data \cite{WA70,UA6}
pin the gluon at  $x \sim 0.4$.
 This information, together with the momentum sum rule, implies
 (i) a decrease of the gluon at medium $x$ ( $\sim 0.1$)  and (ii)
 a steeper  slope of the gluon from $x \sim 10^{-1}$ to $x \sim 10^{-3}$.
 As a consequence  we find (see below) that the gluon obtained from our
 new global analysis  G
 does indeed  change the description of the
$ p \bar p$ collider
data on prompt photon production at small $x_T = 2 p_T^\gamma /\sqrt{s}$.

Another important consequence  of the reduction of the gluon in the
interval $x \sim 0.1 - 0.2$ concerns the prediction for the value of
$\alpha_s$. At these $x$ values the gluon still gives a significant
(positive) contribution to $\partial F_2 / \partial \log Q^2$. Since precise
fixed-target measurements of $F_2$ exist in this $x$ region (see for
example Figs.~6 and 8 of Ref.~\cite{MRSA}), the reduction in the gluon must be
compensated by an increase in the value of $\alpha_s$.  This brings the
deep-inelastic determination of $\alpha_s = 0.114$ more in line with the
world average
value $\alpha_s = 0.117$ \cite{WEBBER}.

Figure~3 compares the A, A$'$ and G  partons
at $Q^2 = 20$~GeV$^2$. We see that the quark distributions
are essentially identical for $x > 0.05$.
 Indeed the A$'$ and G description of the
fixed-target deep-inelastic scattering data, the $W^\pm$ rapidity
asymmetry and the asymmetry of Drell-Yan production  in $pp$ and $pn$
collisions is as good as that obtained by MRS(A), see Ref.~\cite{MRSA}.
However from Fig.~3 we see a sizeable difference  in the A/A$'$ and G
small-$x$ gluon distributions.
Fig.~4 highlights this difference, as well as indicating the $x$ range
of the various experimental constraints  on the gluon.
Comparing the G and A$'$ gluons,
we see (i) that the new HERA measurements of $\partial
F_2/\partial \log Q^2$ (Fig.~2) lead to an enhancement of the gluon for
$x \lapproxeq 0.01$,
(ii) that the fixed-target prompt photon data require the gluon to be unchanged
for $0.35 \lapproxeq x \lapproxeq 0.55$, see Fig.~5, and (iii) that
as a consequence
the G gluon is reduced in the intermediate interval $0.02
\lapproxeq x \lapproxeq 0.2$.
It is also interesting to compare the fraction of the
gluon's momentum in various
intervals of $x$ for $Q^2 = 20$~GeV$^2$ for the two sets of partons:
\begin{center}
\begin{tabular}{|c|c|c|c|c|}
\hline
\rule[-1.2ex]{0mm}{4ex} set & $0 <x <  10^{-3}$ & $10^{-3} < x <  10^{-2}$ &
$10^{-2} < x <  10^{-1}$ & $10^{-1} < x <  1$ \\
\hline
\rule[-1.2ex]{0mm}{4ex} MRS(A$'$) & 3.5\% & 13.4\% & 46.1\% & 37.0\% \\
\rule[-1.2ex]{0mm}{4ex} MRS(G)    & 5.9\% & 16.0\% & 42.3\% & 35.7\% \\
\hline
\end{tabular}
\end{center}

Prompt photon production serves as a strong constraint on the gluon
since it contributes at leading order.
As mentioned above, the previous MRS analyses have used the WA70
fixed-target data \cite{WA70} to pin down the gluon for $x \sim 0.4$.
The procedure adopted was to perform a next-to-leading order
fit with the renormalization and factorization scales optimized
as described in Ref.~\cite{AUR}.
There now exists quite a range of prompt photon data which cover
the interval $0.01 \lapproxeq x \lapproxeq 0.6$.  Following
Ref.~\cite{HUSTON}, we display these data in Fig.~5(a), together
with their next-to-leading order description by the new G and
A$'$ parton sets with the renormalization and factorization scales
set to $\mu = 0.5 p_T^{\gamma }$.  In our plot we show the more recent
ISR data from the R807 collaboration \cite{R807}, rather than the
earlier R806 measurements.  We note that these improved ISR data
have a $p_T^{\gamma}$ distribution whose shape is more in accord
with QCD than those of R806.  From Fig.~5 we see that, in general,
the next-to-leading order QCD description of the prompt photon
data is satisfactory, except that the $p_T^{\gamma}$ distributions
observed by the CDF \cite{CDF}
and UA2 \cite{UA2} collaborations are steeper than is
predicted.  Scale ambiguities could remedy a discrepancy in
normalization but cannot significantly change the shape
of a distribution.  On the other hand the description
of prompt photon production at high energies and small $p_T^{\gamma}$
is complicated by the presence of sizeable photon bremsstrahlung
contributions.  A detailed study \cite{GGRV} of these photonic fragmentation
effects (based on the full next-to-leading order calculations
of Refs.~\cite{NLO}) has been made, and as a result the
QCD predictions are improved.  However the description of the
CDF and UA2 data shown in Fig.~5 already includes these
next-to-leading order effects of photon
fragmentation\footnote{We thank Werner Vogelsang for providing these
predictions} and, although the shape is improved, it is
still not satisfactory.  It was hoped that the steeper MRS(G)
gluon would have further improved the description of these
data.  We see from comparing the two plots of Fig.~5 that
the improvement is, in fact, marginal.  The most
probable explanation \cite{HUSTON}  of the residual discrepancy
in the shape of the CDF and UA2 distributions at small $x_T$
is the presence of intrinsic $k_T$ arising from
non-perturbative or `soft' perturbative effects.
Since no soft resummation calculations are available, we have
modelled these effects by a two-dimensional transverse
momentum convolution of the next-to-leading
order QCD prediction for the $p_T^{\gamma}$
distribution with a Gaussian
intrinsic $k_T$ form, exp$(-k_T^2/k_0^2)$,
and varied the parameter $k_0^2$ to give the best fit to each set of
prompt photon data.  We find that an excellent description of the
shape of the CDF and  UA2  data sets can then be obtained
if $\langle k_T \rangle =\sqrt{\pi}k_0/2$ is taken
 to be 2.4~GeV in both cases.

What other scattering processes could provide information on the
gluon distribution at small $x$?
In the  production of $b \bar b$ pairs in high-energy hadron-hadron collisions
 the gluon enters (quadratically) at leading order via
the subprocess $gg \to b \bar b$. The prediction for the
cross section is however very sensitive
to the choice of $m_b$ and the QCD scale, even at next-to-leading order,
 and reasonable variations
of these parameters can change the normalization by  a factor of 2 \cite{CDFB}.
Furthermore it is difficult experimentally
 to accurately reconstruct the $b$-quark cross section from measurements
 of its decay products ($B$, $J/\psi$, leptons, \ldots).
 We have checked\footnote{We thank Slawek Tkaczyk for performing
 these calculations}
 that  for $p\bar p$ collisions at $\sqrt{s} = 1.8$~TeV
 the differences in the A$'$ and G gluons at small $x$
 only reveal themselves at $p_T(b) \lapproxeq 10\ {\rm GeV}/c$ where,
 unfortunately, there are essentially no CDF or D0 data. At higher
 $p_T(b)$ the shape of the $p_T$ distribution
 measured by CDF \cite{CDFB} is in reasonable
 agreement with the theoretical predictions, although the normalization
 is slightly larger than predicted with `canonical' scale and mass values.

Dijet production in $p\bar p$ collisions can also, in principle,
probe the small $x$ gluon \cite{CDFDIJET,MRSDIJET,GGKDIJET}.
For example, if two jets are produced with equal  transverse
momentum $p_T$ and  pseudorapidity $\eta \gg 1$
then $x_1 \sim 1$, $x_2 \sim 2p_T/\sqrt{s} \; \exp(-\eta) \ll 1$.
A detailed calculation \cite{NIGEL} shows
that at $\sqrt{s} = 1.8$~TeV
the gluon can be probed in this way in the range $0.005 \lapproxeq
 x_g \lapproxeq 0.05$, as indicated in Fig.~4.
However, at present the systematic errors are too large to allow
any definite conclusion to be drawn.

In a large region of phase space, inelastic $J/\psi$
production at HERA should be dominated by the `colour singlet' process
$\gamma g \to J/\psi\; g$ \cite{SINGLET}. Here the gluon is probed at
$ x \approx c M_{\psi}^2 / s_{\gamma p} \ll 1$, where the constant $c$
depends on the cuts used to define the inelastic events \cite{MNSJPSI}.
The theoretical calculation has recently been extended
to next-to-leading order \cite{KRAEMER}. Forthcoming data from
HERA, with for example $50 \lapproxeq \sqrt{s_{\gamma p}} \lapproxeq
150$~GeV and taking $ c = 3.4$ \cite{MNSJPSI},  will allow us to probe
 the interesting small $x \sim 10^{-3} - 10^{-2} $ region,
although it remains to be seen whether the precision will be high enough
to discriminate between  gluons  like those of
A$'$ and G.

Note that although there is a large overlap in the $x$ region
probed by  $p \bar{p}$ dijet and HERA $J/\psi$ production, the
$Q^2$ values are very different, being respectively
$Q^2 \sim 10^3 - 10^4$~GeV$^2$ and $Q^2 \sim 10$~GeV$^2$.
Since the gluon distributions tend to become more similar as $Q^2$
increases, the $J/\psi$ cross section at lower $Q^2$
provides greater discrimination. For example, the ratio
of the G and A$'$  gluons at $x = 10^{-3}$ is 1.71 at
$Q^2 = 5$~GeV$^2$ and only 1.11 at $Q^2 = 5\times 10^3$~GeV$^2$.
The difference at the former $Q^2$ value is indicative of the
difference in the inelastic $J/\psi$ cross section predictions at high
$\sqrt{s_{\gamma p}}$ at HERA.

In conclusion, we have seen that standard sets of next-to-leading
order parton distributions like MRS(A), CTEQ3 and GRV(94) give
a good overall description of the new HERA structure function data.
However at the very smallest $x$ values there is evidence that
a new effect is becoming apparent, namely that
the small-$x$
behaviour  of the gluon and sea-quark distributions  are {\it not}  linked
at $Q^2 = 5$~GeV$^2$. If, as in set MRS(A$'$), we force the distributions
to be the same, then the slope $\partial F_2(x,Q^2) / \partial \log Q^2$
at small $x$ is underestimated.
Our new MRS(G) set accommodates this by allowing the small-$x$
quarks and gluons to have a different shape in $x$.
 In the notation of Eqs.~(1) and (2) we find that
$\lambda_g = 0.30$ and $\lambda_S = 0.07$, though the precise values
are correlated to the parameters $\epsilon_g$ and $\epsilon_S$.
The situation is summarized in Fig.~6, where we show the
theoretical predictions  and experimental measurements of the
structure function $F_2$ at a `typical' small-$x$ value, $x = 0.0004$.
The A$'$ and G curves are labelled with their gluon
and singlet quark effective $\lambda$ values at this $x$ and the starting
value $Q^2 = 4$~GeV$^2$.

 The resulting  MRS(G) `singular' gluon and `flat' sea quark distributions
do not have a ready explanation in terms of either perturbative
or non-perturbative QCD.
On the one hand GLAP evolution from a low scale,
such as performed by GRV \cite{GRV} (see also
\cite{BALL}), develops both a steep
gluon {\it and} a steep sea quark distribution at small $x$.
 The former is evident in
the data, the latter is not. Again Fig.~6 summarizes the situation.
The GRV(94) curve has a similar slope to MRS(G) (both gluons are steep)
but overestimates the data (the GRV quarks are steeper than the G quarks).
 On the other hand it might be argued that the
leading $\log (1/x)$ resummation, encapsulated in the BFKL equation, is more
appropriate at small $x$.  A singular (unintegrated) gluon is obtained, as
required by the data, but again the steepness is fed, in this case via the
$k_T$-factorization theorem, directly into the sea quark
distribution and hence into $F_2$.  Of course the application of
next-to-leading order, leading-twist QCD in the HERA small-$x$ regime
may be too naive. There are some indications that higher-order perturbative
corrections may be non-negligible and that these may affect the
structure function and its $Q^2$ evolution in different ways \cite{ELLIS}.
It is not impossible, for example, that the  $x\to 0$ behaviour
of the quark and gluon distributions {\it are} more similar when
higher-order corrections are fully taken into account. Unfortunately
the theoretical technology necessary to investigate this with any
precision is not yet available.
If, on the other hand, the form of the starting distributions at $Q_0^2$
were primarily of non-perturbative origin, the dominance
of the forward quark-proton and gluon-proton  scattering amplitudes
by pomeron  exchange would again have suggested $\lambda_S \sim \lambda_g$.
In any case, more precise structure function measurements at small $x$
and moderate $Q^2$ will be very useful in shedding further light on this issue.

\section*{Acknowledgements}
We thank Werner Vogelsang for information concerning prompt photon
production and Andreas Vogt for providing the code of the latest
set of GRV partons.

\newpage

\section*{Figure Captions}

\begin{itemize}

\item[{[1]}] The (a) ZEUS \cite{ZEUS} and (b) H1 \cite{H1} measurements of
$F_2(x,Q^2)$ in the small $x$ regime compared
 with the predictions of the  MRS(A)  partons of Ref.~\cite{MRSA}
 (dash-dot curve, shown only for $x \leq 5.6 \times 10^{-4}$),
the  MRS(A$'$)  and MRS(G) partons of this work
(dashed and solid curves respectively),
 and  the GRV(94) partons of Ref.~\cite{GRV} (dotted curve).

\item[{[2]}] (a) The slopes, $b(x)$, determined from linear fits, $F_2 = a +
b \log Q^2$, to the HERA data and (b) the values
of $F_2(x,Q^2 = 5$~GeV$^2$) obtained by
extrapolating the linear fits  back to $Q^2 = 5$~GeV$^2$.  Also shown are the
predictions of  the MRS(A$'$) (dashed curve) and MRS(G) (solid curve)
 partons of this work,
together with the description given by
 the recent GRV(94) partons of Ref.~\cite{GRV} (dotted curve).
  The slopes are calculated at $Q^2 = 20$~GeV$^2$.

\item[{[3]}]  MRS(G) partons compared to MRS(A$'$) and also to the
earlier MRS(A) \cite{MRSA} partons, at
$Q^2 = 20$~GeV$^2$. For clarity,
the gluons are shown for $x < 0.005$ only.

\item[{[4]}] The MRS(A$'$) and MRS(G) gluon distributions
at $Q^2 = 20$~GeV$^2$. Also shown are  the $x$
intervals in which the gluon is constrained by the various sets of data.

\item[{[5]}]  The description of  WA70 \cite{WA70}, UA6 \cite{UA6},
R807 \cite{R807}, UA2 \cite{UA2} and
CDF \cite{CDF}
prompt photon data by the MRS(A$'$) (lower figure) and MRS(G)
(upper figure) partons.
The predictions for the $p \bar p$ collider cross sections
 are calculated using the next-to-leading
order photonic fragmentation program of Ref.~\cite{GGRV},
and for the other cross sections using the next-to-leading
order program of Aurenche {\it et al.} \cite{NLO}.

\item[{[6]}] The description of the ZEUS \cite{ZEUS} and
 H1 \cite{H1} measurements of
$F_2(x,Q^2)$ near $x = 0.0004$  compared  with
the  MRS(A$'$)  and MRS(G) partons of this work
(dashed and solid curves respectively),
 and  the GRV(94) partons of Ref.~\cite{GRV} (dotted curve). The effective
 $\lambda$ values for the gluon and singlet quark distributions
 at this value of $x$ and $Q^2 = 4$~GeV$^2$ are indicated.

\end{itemize}

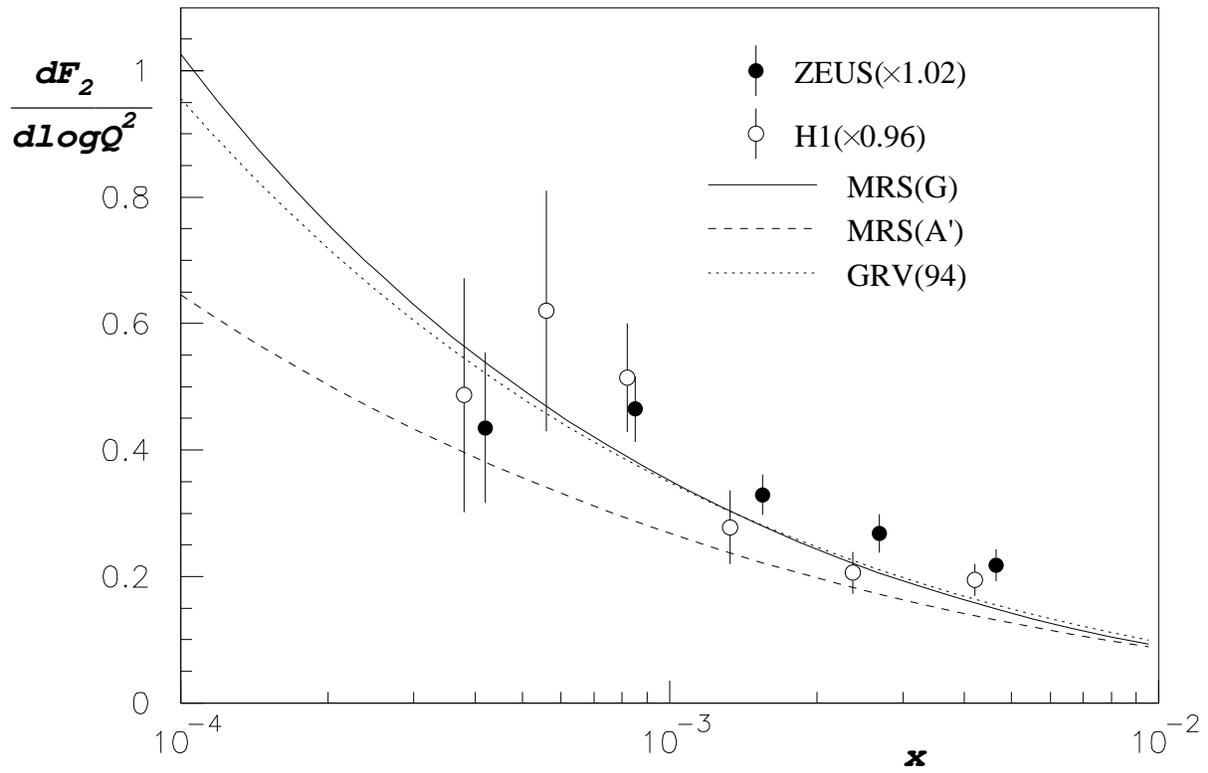
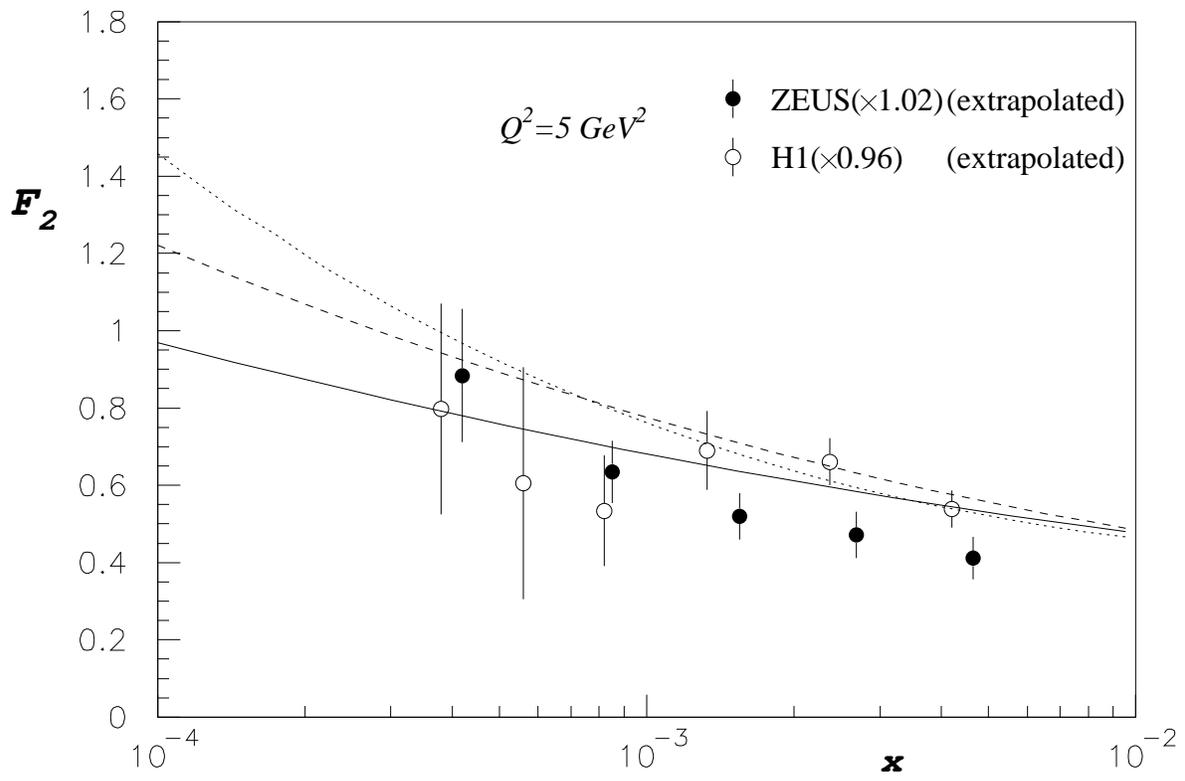
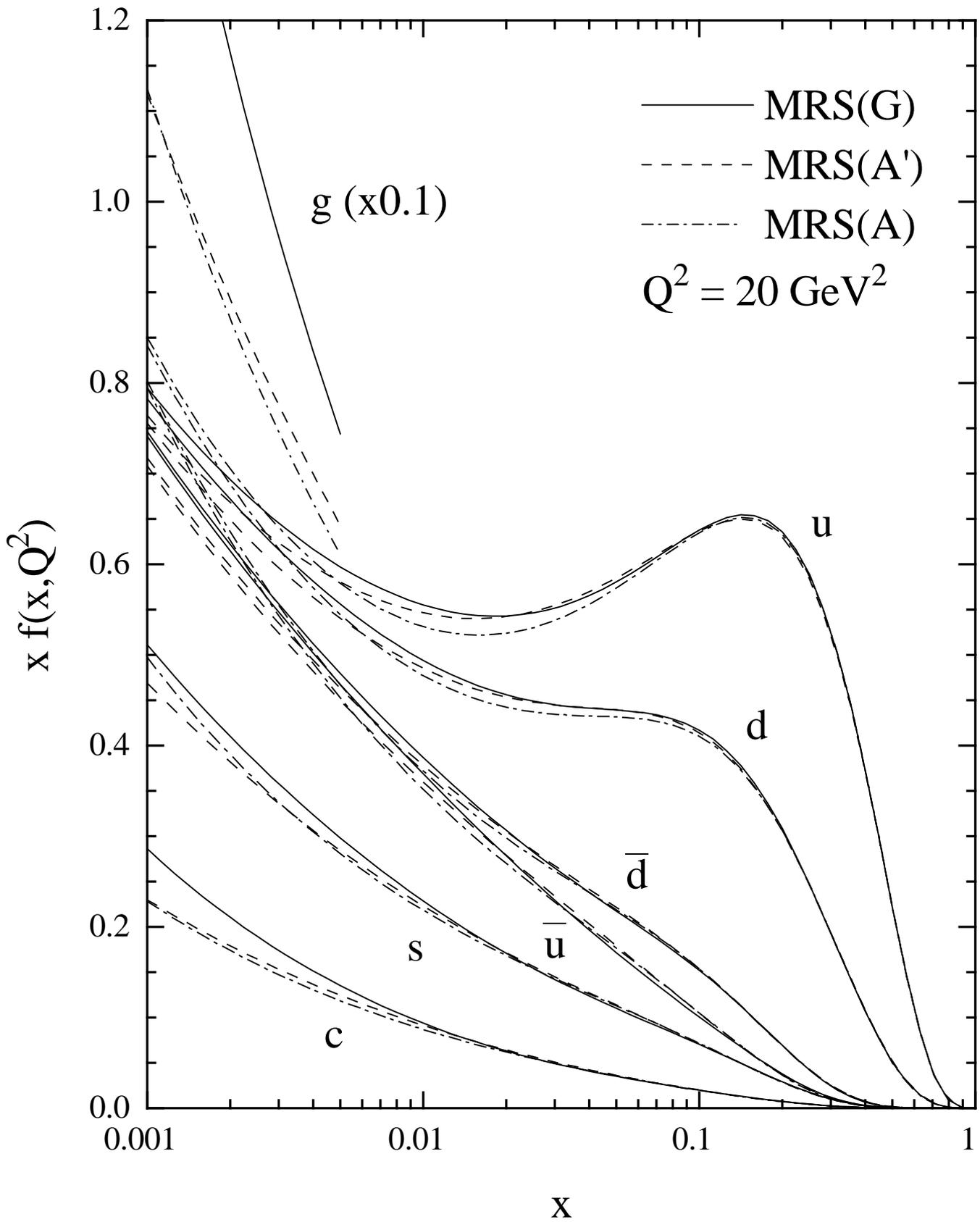
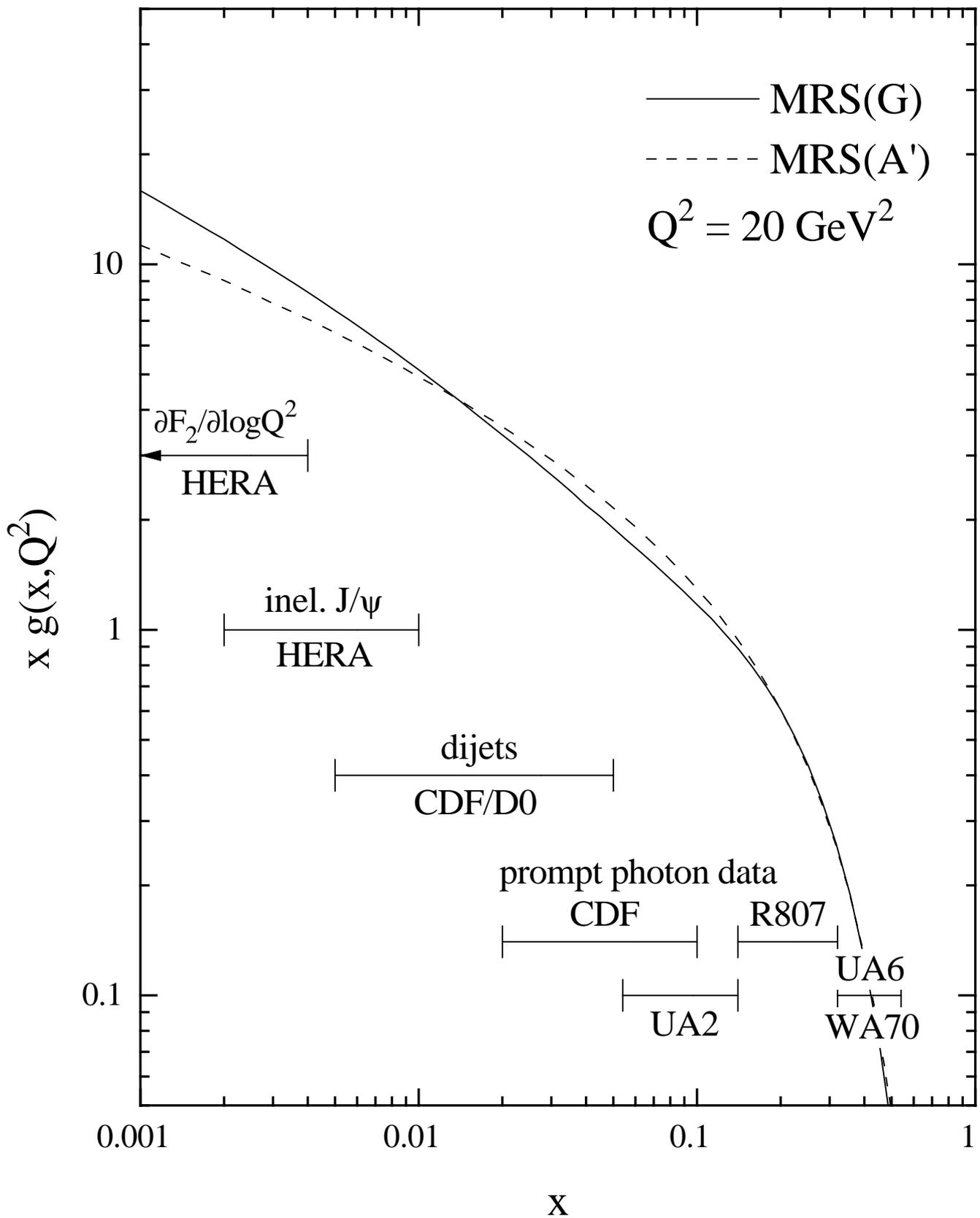
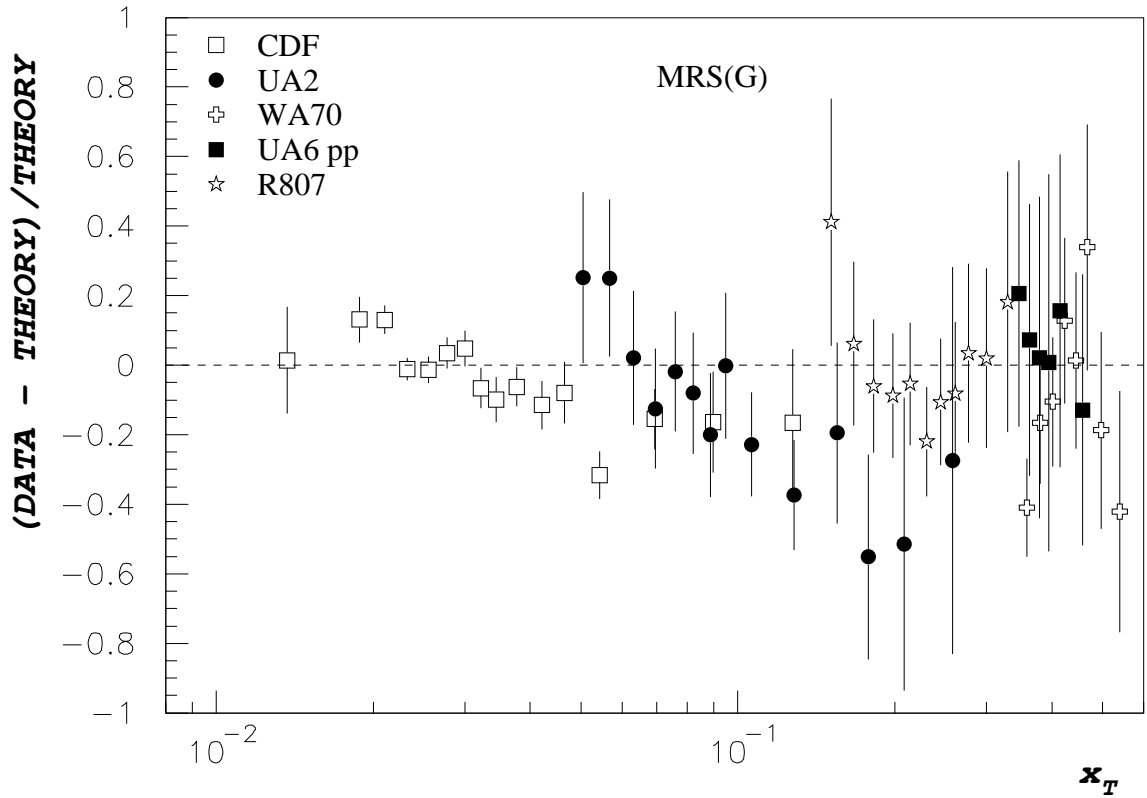
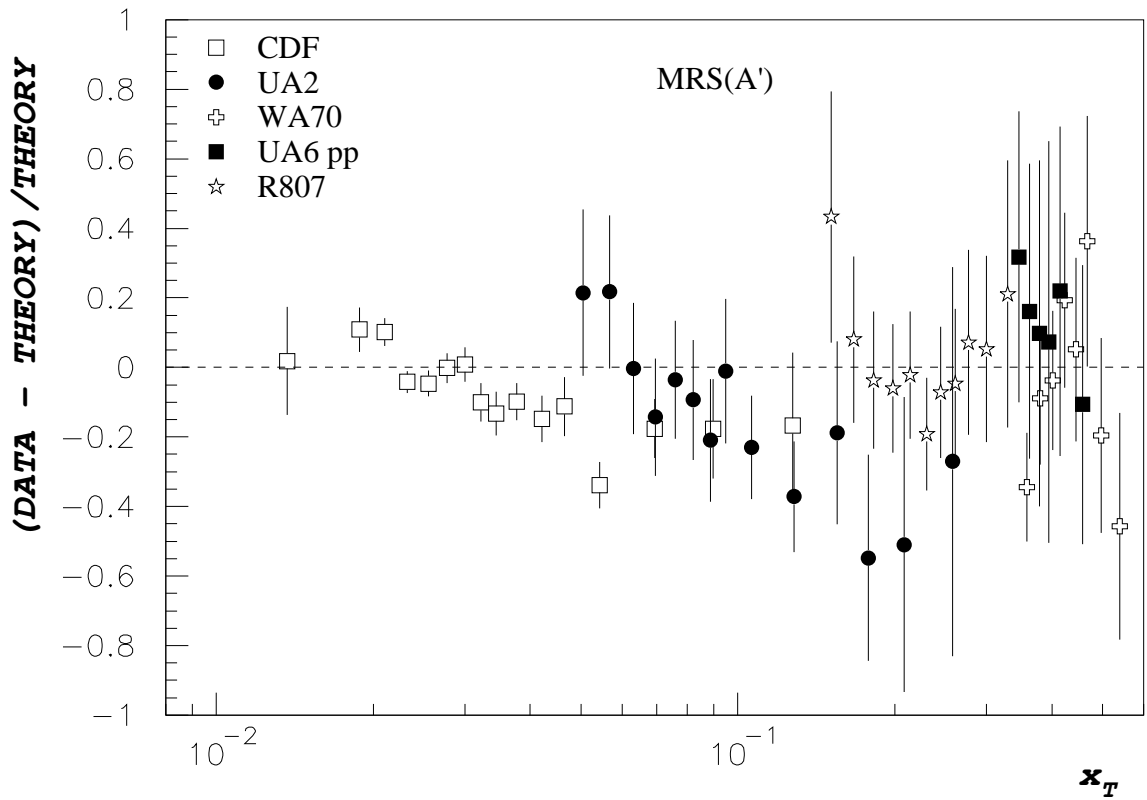
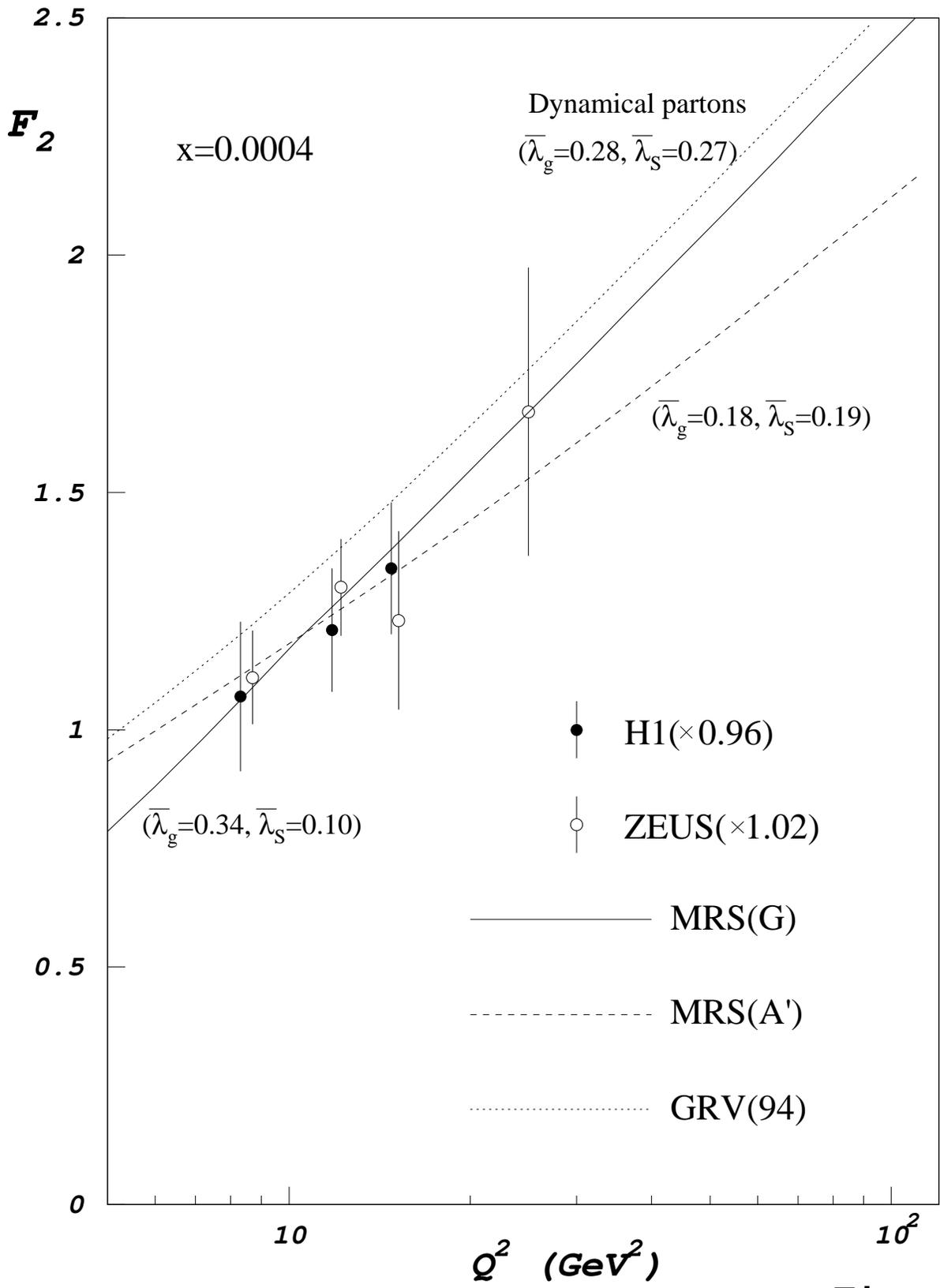
\end{document}